# Symmetry-protected four double-Weyl fermions and their topological phase transitions in nonmagnetic crystals

Yun-Yun Bai[*], Ke-Xin Pang[*], and Yan Gao[†]

*State Key Laboratory of Metastable Materials Science and Technology & Hebei Key Laboratory of Microstructural Material Physics, School of Science, Yanshan University, Qinhuangdao 066004, China*

**Realizing Weyl semimetals (WSMs) with the minimal number of Weyl points (WPs) fundamentally simplifies extracting intrinsic topological responses. While a minimum of four conventional ($|C| = 1$) WPs in nonmagnetic crystals is well-established, the exact symmetry requirements and material realization for the unique configuration of four unconventional double-Weyl points (DWPs, $|C| = 2$) remain unresolved. Here, we establish rigorous crystalline symmetry constraints restricting the existence of exactly four symmetry-protected DWPs to merely 28 space groups in both nonmagnetic spinless and spinful systems. Guided by this classification, we identify an *sp²-sp³* hybridized chiral carbon allotrope, THRLN-C$_{32}$, as an ideal candidate hosting precisely this four-DWP configuration near the Fermi level. These $C_4$-protected DWPs project extended or closed-loop Fermi arcs onto the surface Brillouin zone, providing unambiguous spectroscopic signatures. Furthermore, external strain drives profound topological phase transitions encapsulated in a unified evolution landscape: the pristine four-DWP state dissociates into two exotic three-terminal Weyl complexes, degenerates into eight conventional $|C| = 1$ WPs, or collapses into a trivial insulator. This work provides a definitive theoretical framework for minimal double-WSMs in nonmagnetic spinful systems and introduces an optimal material platform for investigating strain-tunable topological quantum phenomena.**

---

[*]These authors contributed equally to this work.
[†]Corresponding author: yangao9419@ysu.edu.cn

Weyl semimetals (WSMs) represent a fundamental class of topological states of matter, characterized by bulk band crossings that act as monopoles of Berry curvature in momentum space[1-4]. These topological nodes, known as Weyl points (WPs), carry a non-zero chiral topological charge (Chern number, $C$) and drive a host of exotic physical phenomena, including unclosed Fermi arc surface states, the chiral anomaly, and anomalous transport responses[5-10]. The emergence of WPs strictly requires the breaking of either time-reversal ($\mathcal{T}$) and/or spatial inversion ($\mathcal{P}$) symmetry[11, 12]. Beyond conventional WPs with $|C| = 1$, specific crystalline symmetries can protect unconventional WPs with higher topological charges ($|C| = 2,3,4$), which exhibit anisotropic low-energy dispersions and amplified topological responses[13-22]. However, the vast majority of theoretically and experimentally confirmed WSMs possess a large number of WPs distributed at varying energy levels throughout the Brillouin zone (BZ)[23-25]. This multiplicity severely complicates the isolation of topological transport signatures and obscures the fundamental physics of Weyl fermions[25, 26]. Consequently, realizing WSMs with the minimal allowable number of WPs is highly desirable for extracting unambiguous intrinsic topological responses.

The minimum number of WPs in a crystal is strictly governed by symmetry constraints, with $\mathcal{T}$ playing a decisive role. According to the Nielsen-Ninomiya no-go theorem[27, 28], WPs must appear in pairs of opposite chirality to ensure a net-zero topological charge within the BZ. In $\mathcal{T}$-breaking magnetic systems, the theoretical minimum is two WPs, a configuration often referred to as the "hydrogen atom" of WSMs[29-34]. Conversely, in nonmagnetic systems, the constraints are more stringent.

The $\mathcal{T}$ operator maps a WP at momentum *k* to -*k* while preserving its chirality. To satisfy the no-go theorem, an additional pair of WPs with opposite chirality must exist, establishing the theoretical minimum of four conventional ($|C| = 1$) WPs for nonmagnetic WSMs. While a few materials hosting exactly four $|C| = 1$ WPs have been reported[35-41], the realization of the unique configuration for unconventional WPs—specifically, exactly four $|C| = 2$ double-Weyl points (DWPs)—remains largely unexplored. A comprehensive symmetry framework defining the exact crystalline conditions necessary to protect exactly four DWPs in nonmagnetic systems is currently lacking. This gap restricts the systematic discovery of ideal unconventional WSMs.

In this work, we resolve this challenge by establishing a rigorous symmetry analysis framework to identify the definitive search space for nonmagnetic crystals hosting exactly four $|C| = 2$ DWPs. By systematically analyzing the encyclopedia of emergent particles across all 230 space groups (SGs), we isolate exactly 28 SGs capable of accommodating this unique four-DWP configuration in both nonmagnetic spinless and spinful systems. Guided by these exact symmetry constraints, we propose THRLN-$C_{32}$, an $sp^2$-$sp^3$ hybridized chiral three-dimensional carbon allotrope. First-principles calculations and symmetry analysis confirm that THRLN-$C_{32}$ hosts exactly four DWPs located almost precisely at the Fermi level along the high-symmetry Γ-Z path, protected by $C_4$ rotational symmetry. Furthermore, we reveal that these DWPs project distinct, highly extended double or closed-loop Fermi arcs onto the (100) and (110) surfaces, providing stark contrast to conventional open arcs. Crucially, we demonstrate that hydrostatic and uniaxial strains drive a significant sequence of topological phase transitions: the pristine four-DWP state can dissociate into two exotic three-terminal Weyl complexes (each containing one $|C| = 2$ and two $|C| = 1$ WPs), degenerate into four pairs of conventional $|C| = 1$ WPs, or undergo a complete collapse into a trivial insulator. This work maps the exact symmetric boundaries for unique four unconventional fermions and provides an ideal material platform for chiral topological physics.

**Results**

**Symmetry constraints for exactly four $|C| = 2$ double-Weyl points.** We begin by establishing the rigorous algebraic requirements necessary to stabilize exactly four symmetry-protected unconventional double-Weyl points (DWPs, $|C| = 2$) in a nonmagnetic, time-reversal ($\mathcal{T}$) invariant system. Let the four $|C| = 2$ DWPs be located at distinct momenta $\boldsymbol{k}_1$, $\boldsymbol{k}_2$, $\boldsymbol{k}_3$, and $\boldsymbol{k}_4$ ($\boldsymbol{k}_1 \neq \boldsymbol{k}_2 \neq \boldsymbol{k}_3 \neq \boldsymbol{k}_4$), with their corresponding little groups denoted as $G_{\boldsymbol{k}_1}$, $G_{\boldsymbol{k}_2}$, $G_{\boldsymbol{k}_3}$, and $G_{\boldsymbol{k}_4}$. To prevent the proliferation of additional symmetry-related nodes and strictly confine the system to exactly four DWPs, the host space group ($G$) must simultaneously satisfy the following criteria:

(i) The space group $G$ must support Weyl points with topological charge $|C| = 2$.

(ii) If four Weyl points $k_1$, $k_2$, $k_3$, and $k_4$ exist independently, their little groups must identical, $G_{k_1} = G_{k_2} = G_{k_3} = G_{k_4} = G$, thereby preventing the generation of additional symmetry-related WPs.

(iii) If two WPs, for instance $k_1$ and $k_2$, are independent, while the remaining pair $k_3$, and $k_4$ are connected by a space-group operation $\mathcal{O}$ (including proper rotations $\mathcal{R}$, time-reversal symmetry $\mathcal{T}$, or their combination), the SG $G$ admits the coset decomposition $G = G_{k_1} = G_{k_2} = G_{k_3} \cup \mathcal{O} G_{k_3}$, with the little group of $k_4$ given by $G_{k_4} = \mathcal{O} G_{k_3} \mathcal{O}^{-1}$.

(iv) If the four DWPs form two symmetry-related pairs, such that $k_1$ and $k_2$ are related by an operation $\mathcal{O}_1$, while $k_3$ and $k_4$ are related by another operation $\mathcal{O}_2$, the SG can be expressed as $G = G_{k_1} \cup \mathcal{O}_1 G_{k_1} = G_{k_3} \cup \mathcal{O}_2 G_{k_3}$, with the corresponding little groups satisfying $G_{k_2} = \mathcal{O}_1 G_{k_1} \mathcal{O}_1^{-1}$, $G_{k_4} = \mathcal{O}_2 G_{k_3} \mathcal{O}_2^{-1}$, where $\mathcal{O}_1, \mathcal{O}_2 \in \{\mathcal{R}, \mathcal{T}, \mathcal{RT}\}$.

Under these stringent topological charge and algebraic constraints, we perform a systematic survey of all 230 SGs utilizing the encyclopedia of emergent particles[20]. Evaluating both the spinless and spinful limits, we determined that this unique four-DWP configuration is mathematically restricted to exactly 28 SGs. This definitive classification, summarized in Table I, establishes the complete symmetry-allowed landscape for realizing exactly four symmetry-protected $|C| = 2$ DWPs in nonmagnetic crystals.

**Chiral Crystal Structure and Stability of THRLN-C$_{32}$.** Guided by the aforementioned theoretical framework, we seek a pristine material platform capable of satisfying these symmetry constraints. Carbon, with its versatile $sp^2$-$sp^3$ orbital hybridization, provides an optimal basis for constructing complex three-dimensional (3D) chiral networks. Accordingly, we propose THRLN-C$_{32}$ (tetragon–hexagon ring-linked (2,2) nanotube), a novel chiral carbon allotrope. This network crystallizes in two structurally enantiomorphic configurations related by spatial inversion: the left-handed $l$-THRLN-C$_{32}$ adopting the chiral tetragonal SG $P4_322$ (No. 95) [Fig. 1(a)], and the right-handed $r$-THRLN-C$_{32}$ adopting SG $P4_122$ (No. 91) [Fig. 1(b)]. Crucially, both structures belong to the $D_4$ crystallographic point group, perfectly matching our theoretical criteria for hosting four double-Weyl fermions [see Table I].

As illustrated in Fig. 1(e), the optimized lattice parameters are $a = b = 7.32$ Å and $c = 5.00$ Å, with the primitive cell comprising 32 carbon atoms distributed across five inequivalent Wyckoff positions [see Table SI in the Supplemental Material (SM)]. The structural architecture of THRLN-C$_{32}$ is assembled from three fundamental building blocks: one-dimensional (2,2) single-walled carbon nanotubes (CNTs) [Fig. 1(c)], open helical tetragonal ring chains (HTRCs), and helical hexagonal ring chains (HHRCs) [Fig. 1(d)], all seamlessly interconnected via strong covalent bonds. The calculated $sp^2$-$sp^2$ (1.44 Å) and the $sp^2$-$sp^3$ (1.55 Å) bond lengths are in excellent agreement with the corresponding values in graphene (1.42 Å) and diamond (1.54 Å), respectively, reflecting the mixed hybridization character of the framework. Notably, while the (2,2) CNT sublattice remains achiral, the HTRC and HHRC sublattices dictate the macroscopic chirality of the crystal—presenting as strictly left-handed in $l$-THRLN-

$C_{32}$ and right-handed in $r$-THRLN-$C_{32}$. The overarching structure can thus be conceptualized as the lateral coupling of these three distinct structural motifs into a unified, highly rigid helical $sp^2$-$sp^3$ carbon network.

We rigorously verified the physical viability of THRLN-$C_{32}$ through multiple independent stability criteria. Phonon dispersion spectra [Fig. S1 in the SM] reveal a complete absence of imaginary frequencies across the entire BZ, confirming intrinsic dynamical stability. Mechanically, the calculated independent elastic constants strictly satisfy the Born stability criteria[42] for a tetragonal lattice [see Table SII in the SM]. The density of THRLN-$C_{32}$ is 2.38 g/cm$^3$, which is comparable to that of graphite (2.26 g/cm$^3$). Its bulk modulus of 189.2 GPa substantially exceeds that of T-carbon (155.5 GPa), indicating superior compressional stability despite its open porous framework formed by the CNT-HTRCs-HHRCs composite architecture. A comprehensive structural comparison with other relevant carbon allotropes is detailed in Table SIII of the SM.

**Bulk electronic topology and chirality mapping.** As carbon is a light element, spin-orbit coupling (SOC) is negligible, allowing us to treat THRLN-$C_{32}$ as a spinless system. The calculated bulk electronic band structures for both enantiomers [Figs. 2(a) and 2(b)] reveal that the conduction and valence bands (specifically the 64th and 65th bands) intersect exclusively along the high-symmetry Γ-Z line. This intersection forms twofold degenerate crossing points, denoted as $W_1$ and $W_2$, situated almost exactly at the Fermi level ($E_F$). In $l$-THRLN-$C_{32}$, $W_1$ and $W_2$ are located at +3.65 meV and -9.14 meV relative to $E_F$, respectively. The projected density of states (PDOS) exhibits a pronounced dip approaching zero near $E_F$ [Figs. 2(c) and 2(d)], confirming a quintessential semimetallic character dominated by the *p*-orbital contributions from the $sp^2$-hybridized carbon atoms.

Detailed dispersion analysis [Figs. 2(i)-2(l)] reveals that both $W_1$ and $W_2$ exhibit linear band crossings along the $k_z$ direction, yet display distinctly quadratic dispersions within the $k_x$-$k_y$ plane. This highly anisotropic dispersion profile is the hallmark of unconventional ($|C| = 2$) double-Weyl fermions. By computing the evolution of Wannier charge centers (WCCs) on a sphere enclosing each individual node, we quantitatively determine that $W_1$ and $W_2$ in *l*-THRLN-$C_{32}$ carry chiral topological charges of $C = +2$ and $C = -2$, respectively [Figs. 2(e) and 2(f)]. Notably, the macroscopic structural chirality dictates the momentum-space topological distribution: applying spatial inversion to map *l*-THRLN-$C_{32}$ onto its right-handed enantiomer strictly reverses the topological charges of the corresponding DWPs [Figs. 2(g) and 2(h)]. An exhaustive scan of the entire BZ confirms the existence of exactly four DWPs—two pairs related by time-reversal symmetry $\mathcal{T}$ and protected by the $C_4$ crystalline rotation—establishing THRLN-$C_{32}$ as the ideal realization of the four-DWP configuration.

**Effective $\boldsymbol{k \cdot p}$ Models and symmetry protection.** To establish the topological protection of the DWPs on symmetry grounds and to elucidate the origin of the anisotropic dispersion, we construct the $\boldsymbol{k \cdot p}$ effective Hamiltonian at each DWP. Because all four DWPs ($W_1$ and $W_2$ in both *l*- and *r*-THRLN-$C_{32}$) strictly reside on the high-symmetry Γ-Z line, their low-energy physics is universally governed by the same $C_4$ little group and originates from the intersection of the identical one-dimensional irreducible representations (IRRs), $\Lambda_1$ and $\Lambda_2$ [see Figs. 2(i, k, m, o)]. Consequently, a single unified effective model characterizes all four topological nodes in both *l*- and *r*-THRLN-C$_{32}$, with the only distinction being the fractional momentum coordinate $w$ along the Γ-Z path.

Taking the W$_1$ node of the right-handed enantiomer *r*-THRLN-C$_{32}$ (SG $P4_122$ (No. 91)) as a representative case, the essential symmetry generator is the screw rotation $\tilde{C}_{4z} = \{C_{4,001}^+|00\frac{1}{4}\}$. Utilizing the eigenstates of $\Lambda_1$ and $\Lambda_2$ as the basis, the representation matrix for this symmetry generator takes the diagonal form:

$$D\{\tilde{C}_{4z}\} = \begin{bmatrix} e^{i\pi w/2} & 0 \\ 0 & e^{i\pi(2+w)/2} \end{bmatrix} \tag{1}$$

where $w$ parameterizes the precise location of the node along the Γ-Z path ($w = \pm 0.145$ for W$_1$ in *r*-THRLN-C$_{32}$). The symmetry operation $\tilde{C}_{4z}$ strictly constraints the Hamiltonian via the relation:

$$D(\tilde{C}_{4z})H(\boldsymbol{k})D^{-1}(\tilde{C}_{4z}) = H(\tilde{C}_{4z}\boldsymbol{k}). \tag{2}$$

By expanding the local momentum $k$ up to the quadratic order ($k^2$), the Hamiltonian satisfying this symmetry constraint is obtained as:

$$H_{\text{W1}}(\boldsymbol{k}) = \varepsilon(\boldsymbol{k}) + d(\boldsymbol{k})\sigma_z + [(\alpha k_+^2 + \beta k_-^2)\sigma_+ + \text{H.c.}], \tag{3}$$

where $\varepsilon(\boldsymbol{k}) = \epsilon_0 + \epsilon_1 k_z + \epsilon_2(k_x^2 + k_y^2) + \epsilon_3 k_z^2$, $d(\boldsymbol{k}) = \delta_1 k_z + \delta_2(k_x^2 + k_y^2) + \delta_3 k_z^2$, $k_\pm = k_x \pm i k_y$, $\sigma_\pm = (\sigma_x \pm i\sigma_y)/2$, and $\sigma_i$ ($i = x, y, z$) are there Pauli matrices. The parameters $\epsilon_i$ and $\delta_i$ ($i = 0,1,2,3$) are real, whereas $\alpha$ and $\beta$ are complex coefficients determined by fitting to the underlying electronic band structure.

Equation (3) makes the topological character of the node explicit. The diagonal terms $\varepsilon(\boldsymbol{k})$ and $d(\boldsymbol{k})$ contain linear leading contributions in $k_z$ (namely $\epsilon_1 k_z$ and $\delta_1 k_z$), which generate a linear dispersion along the $C_4$ axis. Concurrently, the off-diagonal terms couple exclusively via $k_\pm^2$, dictating a purely quadratic dispersion within the transverse $k_x$–$k_y$ plane. This anisotropic linear–quadratic dispersion uniquely identifies the crossing as a DWP carrying a topological charge of $|C| = 2$. By fitting the parameters of $H_{W1}(\boldsymbol{k})$ to the DFT bands of $r$-THRLN-C$_{32}$, the resulting analytical dispersion demonstrates excellent agreement with the DFT-derived bands in the vicinity of the W$_1$ and W$_2$ nodes [see Fig. S2 in the SM], unambiguously validating our effective topological description for THRLN-C$_{32}$.

**Topological surface states and closed-loop Fermi arcs.** Figures 3(a) and 3(b) display the surface band structures of $l$-THRLN-C$_{32}$ and $r$-THRLN-C$_{32}$ on the (100) surface. Dictated by bulk-boundary correspondence, each DWP projected onto this surface must be connected to two Fermi arcs, as each node carries a topological charge of $|C| = 2$. The constant-energy contours at +4 meV and −3 meV for $l$-THRLN-C$_{32}$ [Figs. 3(c) and 3(d)] reveal that the DWPs projected onto $\bar{Z}$-$\bar{\Gamma}$ connect to DWPs of opposite chirality on $-\bar{Z}$-$\bar{\Gamma}$. This unique routing forms longitudinally connected, closed-loop Fermi arcs—a boundary topology qualitatively distinct from the open Fermi arcs typical of conventional WSMs. In $r$-THRLN-C$_{32}$, the constant-energy contour at +1 meV [Fig. 3(e)] demonstrates that the Fermi arcs extend laterally and traverse the first BZ boundary (marked by yellow dashed lines), reflecting a different DWP pairing channel driven by the inversion of structural chirality. The profound diversity of Fermi arc geometries between the two enantiomers provides a direct experimental signature of the non-trivial pairing between the four DWP projections.

Figures 3(f) and 3(g) present the (110) surface band structures of *l*-THRLN-$C_{32}$ and *r*-THRLN-$C_{32}$. In stark contrast to the (100) surface, the (110) projection enforces a nearest-neighbor DWP pairing motif: projected WPs of opposite chirality that reside in closest proximity combine to form compact, closed Fermi rings. Each ring is constituted by a pair of Fermi arcs connecting the projections of a chirality-opposite pair. The resulting constant-energy contours at +4 meV and +1 meV [Figs. 3(h) and 3(i)] display two independent, closed Fermi loops—a topological novelty arising directly from the coincident projection geometry of the four DWPs on this specific termination.

Crucially, the extreme energetic proximity of all four DWPs to $E_F$ (within $\pm 12$ meV) ensures that these exotic surface features are immediately accessible to angle-resolved photoemission spectroscopy (ARPES) measurements without the necessity for external gating or chemical doping. This provides unique background-free experimental fingerprints of the underlying unconventional WSM state.

**Strain-driven topological phase transitions and the complete evolution landscape.** To probe the topological robustness and dynamic tunability of THRLN-$C_{32}$, we simulate the electronic response of *r*-THRLN-$C_{32}$ under both hydrostatic and uniaxial strain. The underlying mechanism governing the nodal topology along the high-symmetry Γ-Z path is fundamentally dictated by the $C_4$ rotational symmetry. Specifically, the two crossing bands, denoted as $\Lambda_1$ and $\Lambda_2$ [Fig. 4], correspond to distinct one-dimensional IRRs of the $C_4$ rotation with eigenvalues $\lambda_1$ and $\lambda_2$. The topological charge of the crossing is rigorously determined by their eigenvalue ratio[13], $\eta = \lambda_1/\lambda_2$: a ratio of $\eta = -1$ enforces a DWP with $|C| = 2$, whereas $\eta = \pm i$ yields a conventional Weyl point with $|C| = 1$.

Under positive hydrostatic pressure (from 0 to 25 GPa), the critical $\eta = -1$ eigenvalue ratio is strictly preserved [Fig. 4(a)], forcing the DWPs of opposite chirality to mutually approach one another along the $k_z$ axis. At 25 GPa, these nodes annihilate, opening a global continuous bandgap and driving a topological phase transition from the unique four-DWP WSM phase into a topologically trivial insulator [Fig. 4(a)]. The persistence of the DWPs up to such high pressures demonstrates the exceptional structural and topological robustness of the lattice against compressive perturbations.

Remarkably, applying negative hydrostatic pressure (-7 GPa) or uniaxial tensile strain along the c-axis (+6%) preserves the $C_4$ rotational symmetry and thus maintains the critical $\eta = -1$ invariant along the rotational axis [Figs. 4(a) and 4(b)], triggering a significant topological reconfiguration [Fig. 4(c)]. Because the structural symmetries remain intact, the original pairs of identical-chirality DWPs located proximal to the Γ point cannot simply annihilate to satisfy the no-go theorem. Instead, the bands undergo a symmetry-preserving topological node dissociation. Taking the strained *r*-THRLN-$C_{32}$ at -7 GPa as the prototypical example [Fig. 5], band crossing analysis reveals that the two original $C = -2$ DWPs proximal to Γ decompose within the $k_x$-$k_y$ plane into four identically chiral conventional WPs with $C = -1$ [Fig. 5(h)]. Specifically, along the high-symmetry Γ-Z and Γ-X paths, the bands intersect to form newly distributed nodes, denoted as $W_2'$ and $W_1'$, respectively.

Detailed dispersion analysis [Figs. 5(d) and 5(e)] confirms that $W_2'$ retains the defining linear-quadratic mixed character, bearing a topological charge of $C = +2$ [Fig. 5(g)]. In stark contrast, $W_1'$ exhibits purely linear dispersion across all three momentum directions [Figs. 5(b) and 5(c)], characteristic of a conventional WP with $C = -1$ [Fig. 5(f)]. An exhaustive scan of the strained BZ identifies a total of six WPs: two $W_2'$ nodes located on the $k_z$ axis (related by $\mathcal{T}$), and four $W_1'$ nodes residing on the $k_z = 0$ plane [Fig. 5(h)]. This indicates that the strain induces the dissociation of the original four DWP configuration into two sets of novel three-terminal Weyl complexes (TTWCs), each comprising one $C = +2$ node flanked by two $C = -1$ nodes [Fig. 4(c)]. This dissociation rigorously conserves the global topological charge ($2 \times \{+2, -1, -1\} = 0$). Consequently, the surface boundary signatures also evolve dramatically, projecting superimposed double Fermi arcs on the (110) surface [Figs. 5(i)-5(l)] and generating striking quadruple spiral Fermi arcs on the (001) surface [Fig. S3 in the SM]. A qualitatively identical topological dissociation into TTWCs is observed under a +6% uniaxial strain along the *c*-axis [see Fig. S4 in the SM].

Furthermore, the topological fate of the system is strictly dictated by the $C_4$ symmetry. If this rotational symmetry is explicitly broken, such as by applying a +2% uniaxial tensile strain along the *a*/*b*-axis [Fig. 6], the $\eta = -1$ topological protection is instantly lifted. Under these conditions, the four DWPs immediately degenerate into four pairs (eight in total) of conventional $|C| = 1$ WPs. Taking the symmetry-broken *r*-THRLN-C$_{32}$ (+2% *a*-axis strain) as an example [Fig. 6], the nodes shift to generic $k$ points, forming two inequivalent $W_1''$ and $W_2''$. Both sets of nodes exhibit strictly linear dispersion across all three momentum directions [Figs. 6(b)-6(e)] and carry topological charges of -1 and +1, respectively [Figs. 6(f) and 6(g)]. An exhaustive BZ scan confirms the presence of eight conventional WPs located off the high-symmetry axes [Fig. 6(h)]. Consequently, the surface boundary signatures revert to standard, unclosed open Fermi arcs connecting opposite-chirality projections on the (110) surface [Figs. 6(i)-6(l)], presenting a stark topological contrast to the closed loops of the pristine phase. Similar symmetry-breaking degenerations are also induced by asymmetric biaxial strain [Fig. S5(b) in the SM].

Collectively, these strain-induced transitions can be encapsulated in a unified topological phase diagram [Fig. 4(c)]. This global landscape clearly illustrates that the pristine four-DWP state serves as a versatile topological mother phase: it fractionates into an exotic TTWC phase under $C_4$-preserving perturbations, degrades into a conventional WSM phase under $C_4$-breaking fields, and ultimately annihilates into a trivial insulator under extreme compression. These results establish mechanical strain as a highly versatile tuning parameter for dynamically reconfiguring the complete Weyl fermion configuration.

**Discussion**

The minimal number of nodes in nonmagnetic double-Weyl semimetals is fundamentally determined by the position of the WPs relative to time-reversal invariant momenta (TRIMs). Two distinct scenarios arise for unconventional $|C| = 2$ WPs in such systems. First, when both DWPs are located at a TRIM, each node is invariant under $\mathcal{T}$ and thus coincides with its own time-reversed partner. In this case, a minimal configuration consisting of two nodes is, in principle, allowed, but only in the spinless limit where SOC is absent[22, 43]. In contrast, when one DWP resides at a generic momentum away from any TRIM, as realized in THRLN-$C_{32}$, $\mathcal{T}$ maps a node at $k$ to another node of identical chirality at $-k$. The no-go theorem constraint then enforces the presence of pairs of opposite chirality, requiring at least two such pairs and yielding a minimal total of four DWPs. THRLN-$C_{32}$ perfectly instantiates this constrained scenario in the non-TRIM class.

Beyond its bulk topology, the structural architecture of THRLN-$C_{32}$ provides a unique platform for anisotropic quantum transport. The composite framework of 1D (2,2) CNTs and helical chains creates strongly anisotropic conduction channels parallel to the $C_4$ rotation axis. Consequently, charge carriers funneled along the $c$ axis experience the linearly dispersing DWP bands, while in-plane transport is governed by the quadratic dispersions. This coupling of structural anisotropy to the unconventional Weyl dispersion offers an ideal solid-state laboratory for isolating direction-dependent transport coefficients, such as the anomalous Hall conductivity[44] and the chiral magnetic effect[45].

Furthermore, the role of macroscopic structural chirality warrants explicit attention. The left-handed ($P4_322$) and right-handed ($P4_122$) enantiomers are related by spatial inversion, which strictly reverses the sign of the topological charges at every WP. This direct correlation implies that chiral photocurrent responses, topological optical effects directly proportional to the WP Chern numbers, will exhibit opposite signs between the two enantiomers. This enables a chiroptical discrimination of the enantiomeric pair directly tied to the bulk topology, offering pristine signatures for second-harmonic generation (SHG) and other Berry-curvature-driven nonlinear optical responses[46, 47].

Finally, we emphasize that THRLN-$C_{32}$ serves merely as a representative example to elucidate the rich topological physics and the complete sequence of phase transitions inherent to this symmetry framework. Importantly, the crystalline classification we established, confining the exactly four $|C| = 2$ configurations to 28 SGs, is broadly applicable beyond electronic materials. This framework applies equally to bosonic quasiparticles, providing a systematic route to predicting and realizing four $|C| = 2$ topological nodes in phononic[48] and photonic[49] systems. Moreover, the dynamic tunability of these phases unveils the fundamental fractionalization pathways of higher-fold topological charges. The strain-induced evolution from double-WSM to TTWCs, and ultimately to conventional WSMs, demonstrates a strict hierarchical structure of topological stability dictated entirely by the underlying $C_4$ symmetry.

In summary, we have established the definitive symmetry criteria for realizing exactly four unconventional DWPs in nonmagnetic crystals, mathematically restricting this fundamental topological phase to 28 specific SGs. By identifying the chiral carbon network THRLN-$C_{32}$ as the prototypical realization, we demonstrate that these symmetry-protected nodes, which reside precisely at the Fermi energy, generate distinct bulk-boundary correspondence manifesting as highly anisotropic, closed-loop Fermi arcs. Crucially, the direct mapping between macroscopic structural chirality and the sign of the momentum-space topological charges provides a robust and reversible tuning knob. Furthermore, our discovery of dynamic topological fractionalization, where symmetry-preserving strain dissociates the original state into two exotic three-terminal Weyl complexes while symmetry-breaking strain degenerates it into four pairs of conventional $|C| = 1$ WPs, elegantly maps the hierarchical stability of these higher-fold fermions. Ultimately, this work delivers both a universal theoretical framework for exactly four symmetry-protected double-Weyl topological quasiparticles and an ideal physical platform for engineering dynamic, chirality-driven quantum phenomena.

**Methods**

**Computational methods:** First-principles calculations of the electronic structure and phonon dispersion spectra for THRLN-$C_{32}$ were performed within the frameworks of density functional theory (DFT)[50, 51] and density functional perturbation theory (DFPT)[52], respectively, as implemented in the Quantum ESPRESSO package[53]. The electron-ion interactions were described using optimized norm-conserving Vanderbilt (ONCV) pseudopotentials[54], and the exchange-correlation functional was treated via the generalized gradient approximation (GGA) parameterized by Perdew, Burke, and Ernzerhof (PBE)[55]. The electronic wave functions and charge density were expanded using a plane-wave basis set with strict kinetic energy cutoffs of 80 Ry and 450 Ry, respectively. To ensure high structural precision, the lattice parameters and internal atomic coordinates were fully relaxed utilizing the Broyden–Fletcher–Goldfarb–Shanno (BFGS) algorithm until the residual Hellmann-Feynman forces on all constituent atoms fell below $10^{-3}$ eV/Å, alongside a total energy convergence criterion of $10^{-6}$ eV. The charge density was integrated on a Γ-centered $12 \times 12 \times 16$ $k$-point mesh[56]. The mechanical stability was evaluated by deriving the independent elastic constants $C_{ij}$ through the energy-strain approach[57]. Furthermore, for the rigorous investigation of topological signatures, an effective tight-binding (TB) Hamiltonian was constructed by projecting the DFT Bloch states onto maximally localized Wannier functions (MLWFs) comprising the carbon $s$ and $p$ orbitals via the Wannier90 code[58]. The subsequent extraction of surface state spectral functions, constant-energy Fermi arcs, and the evolution of Wannier charge centers (WCCs) for precise topological charge determination were executed utilizing the WannierTools Packages[59].

## Data availability

The data supporting the findings are displayed in the main text and the Supplementary Information. All raw data are available from the corresponding authors upon request.

## Acknowledgements:

We wish to thank Weikang Wu for helpful discussions. This work was supported by the National Natural Science Foundation of China (Grants No. 12304202), Hebei Natural Science Foundation (Grant No. A2023203007), Science Research Project of Hebei Education Department (Grant No. BJK2024085).

## Author contributions:

Y.-Y. Bai proposed the THRLN-$C_{32}$ structure and performed the corresponding DFT and DFPT calculations. K.-X. Pang carried out the symmetry analysis. Y. Gao conceived the original ideas. Y.-Y. Bai and K.-X. Pang analyzed the results. Y. Gao writing-reviewing, conceptualization, supervision, and project administration. All authors discussed the results and commented on the manuscript at all stages.

## Competing interests

The authors declare no competing financial interests.

# Figures and Tables

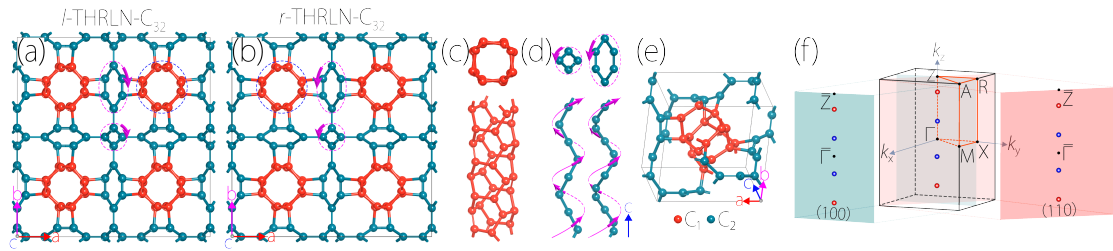

**Figure 1 | Crystal structure and macroscopic chirality of THRLN-C$_{32}$. a**, **b**, Top views of the $2 \times 2 \times 4$ supercells for the enantiomorphic pair: (a) left-handed $l$-THRLN-C$_{32}$ (SG $P4_322$, No. 95) and (b) right-handed $r$-THRLN-C$_{32}$ (SG $P4_122$, No. 91). The fundamental building blocks are highlighted by dashed outlines: (c) one-dimensional achiral (2,2) carbon nanotubes (red), and (d) open helical tetragonal and hexagonal carbon ring chains (cyan). **e**, Perspective view of the primitive cell of $r$-THRLN-C$_{32}$. **f**, Momentum-space distribution of the exactly four DWPs (red and blue spheres denote opposite topological charges) within the bulk Brillouin zone (BZ), alongside their corresponding projection geometries onto the (100) and (110) surfaces.

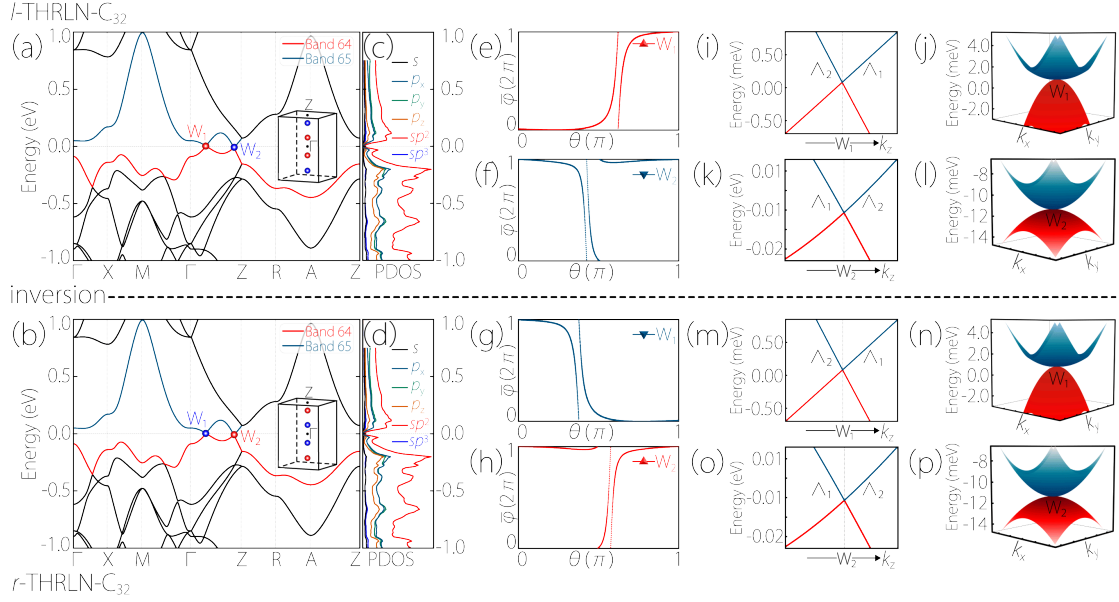

**Figure 2 | Bulk electronic structure and momentum-space topological charge distribution.**
**a**, **b**, Calculated bulk electronic band structures for (a) *l*-THRLN-$C_{32}$ and (b) *r*-THRLN-$C_{32}$ along high-symmetry paths. The crossing of the 64th and 65th bands (red and blue lines) generates exactly four symmetry-protected DWPs exclusively along the Γ-Z invariant axis ($W_1$ and $W_2$). Insets illustrate the spatial distribution of the DWPs within the BZ. Projected density of states (PDOS) for **c**, *l*-THRLN-$C_{32}$ and **d**, *r*-THRLN-$C_{32}$. **e-h**, Evolution of the Wannier charge centers (WCCs) computed on a closed sphere enclosing individual DWPs: $W_1$ and $W_2$ for (**e**, **f**) *l*-THRLN-$C_{32}$ and (**g**, **h**) *r*-THRLN-$C_{32}$. The topological charges strictly reverse sign between the two enantiomers. **i-p**, Detailed energy dispersion profiles around the $W_1$ and $W_2$ nodes for (**i-l**) *l*-THRLN-$C_{32}$ and (**m-p**) *r*-THRLN-$C_{32}$. All nodes exhibit a linear dispersion along the $k_z$ axis and a quadratic dispersion within the $k_x$-$k_y$ plane, rigorously defining the $|C| = 2$ unconventional Weyl fermions. The IRRs ($\Lambda_1$, $\Lambda_2$) associated with the crossing bands are explicitly labeled.

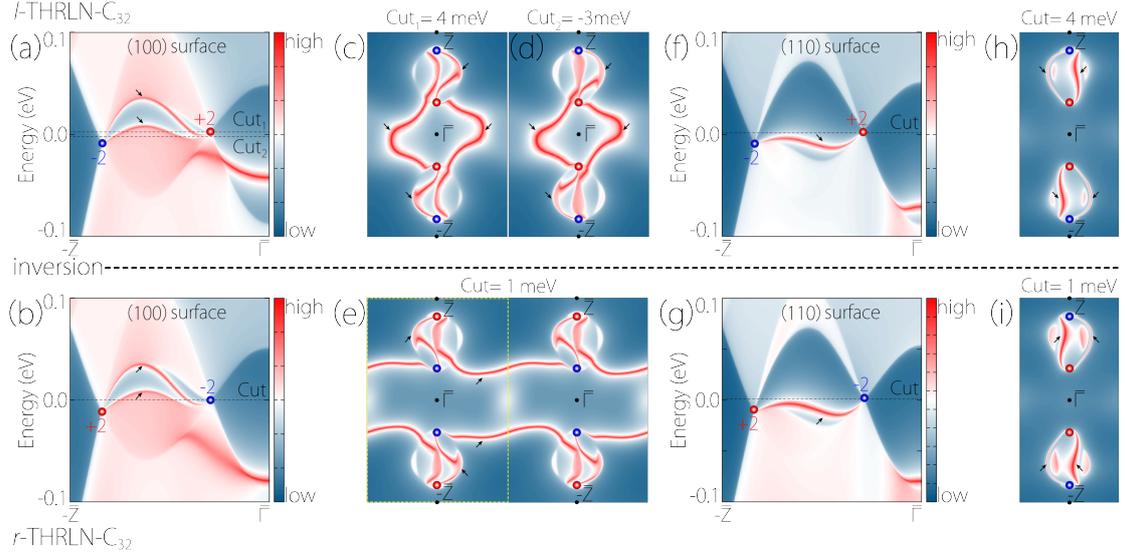

**Figure 3 | Topological surface states and distinct closed-loop Fermi arcs. a**, **b**, Calculated surface spectral functions projected onto the (100) surface for (a) *l*-THRLN-$C_{32}$ and (b) *r*-THRLN-$C_{32}$. Dashed lines indicate the energy positions of the constant-energy contours shown in (**c–e**), and arrows show Fermi arc positions. **c**, **d**, constant-energy contours for *l*-THRLN-$C_{32}$ at +4 meV and -3 meV on the (100) surface. The $|C| = 2$ surface projections emit two Fermi arcs (indicated by black arrows) that connect with opposite-chirality nodes, forming closed-loop Fermi geometries. **e**, Isoenergy contour at +1 meV for *r*-THRLN-$C_{32}$ on the (100) surface, revealing lateral Fermi arc connectivity traversing the BZ boundary (yellow dashed lines) due to structural chirality inversion. **f**, **g**, Surface spectral functions on the (110) boundary for (f) *l*-THRLN-$C_{32}$ and (g) *r*-THRLN-$C_{32}$. **h**, **i**, Corresponding isoenergy contours at +4 meV and +1 meV on the (110) surface. The nearest-neighbor pairing motif forces the surface states to form two distinct, compact closed Fermi rings.

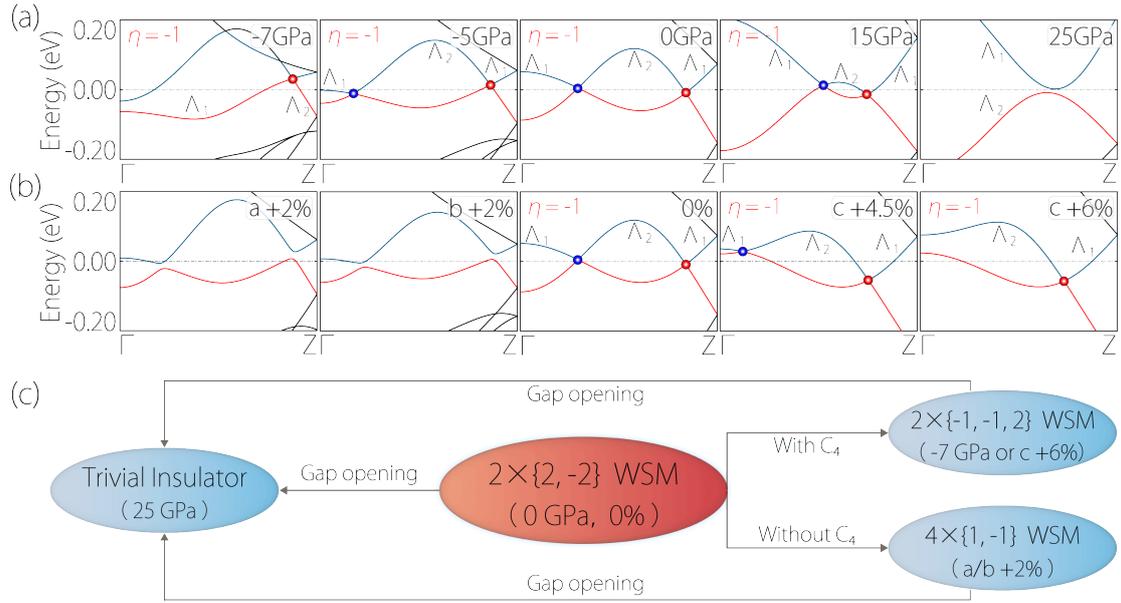

**Figure 4 | Strain-driven topological phase transitions and unified topological phase diagram. a**, **b**, Evolution of the electronic band structure of *r*-THRLN-C$_{32}$ under (a) hydrostatic pressure and (b) uniaxial strain along the *a*, *b*, and *c* axes. The IRRs ($\Lambda_1$, $\Lambda_2$) of the crossing bands are explicitly labeled, with their eigenvalue ratio $\eta = -1$ confirming the $|C| = 2$ DWPs. **c**, Unified topological phase diagram illustrating the dynamic evolution pathways of *r*-THRLN-C$_{32}$ under various mechanical strains. The pristine phase (red) undergoes symmetry-preserving transitions (-7 GPa or *c* +6%) into a phase hosting two sets of TTWCs, experiences symmetry-breaking degenerations (*a*/*b* +2%) into a conventional WSM phase with four pairs of $|C| = 1$ WPs, and ultimately annihilates into a trivial insulator under extreme compression.

**27 / 30**

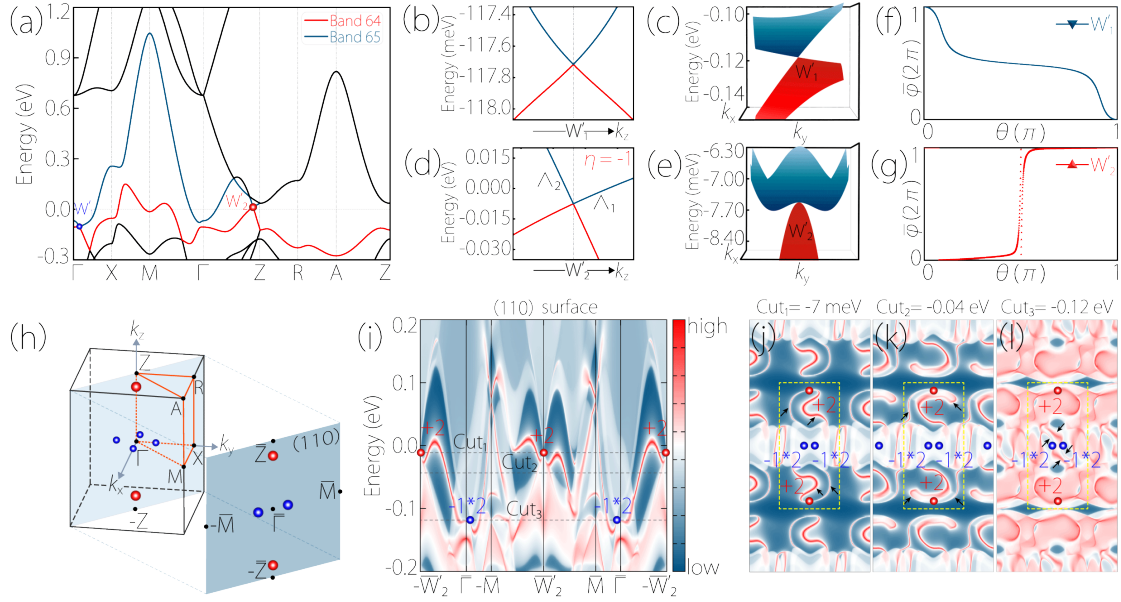

**Figure 5 | Topological node dissociation into two TTWCs under a -7 GPa hydrostatic pressure. a**, Bulk band structure of the strained *r*-THRLN-C$_{32}$ along high-symmetry paths. **b, c**, Purely linear energy dispersions of W$'_1$ along three momentum directions ($k_x$, $k_y$, $k_z$). **d, e**, Energy dispersions of W$'_2$, exhibiting quadratic behavior in the $k_x$-$k_y$ plane and linear behavior along the $k_z$ axis. The IRRs ($\Lambda_1$, $\Lambda_2$) and the eigenvalue ratio $\eta = -1$ are explicitly marked. **f, g**, Evolution of the Wannier charge centers (WCCs) on a closed sphere enclosing W$'_1$ and W$'_2$, confirming topological charges of -1 and +2, respectively. **h**, Reconfigured spatial distribution of the six WPs in the BZ and their corresponding projections onto the (110) surface. **i**, Surface spectral function calculated on the (110) boundary. **j-l**, Isoenergy contours at -7 meV, -0.04 eV, and -0.12 eV, respectively. The yellow dashed lines represent the first BZ boundaries, and the black arrows indicate the evolution of the superimposed double Fermi arcs connecting projections of opposite chirality.

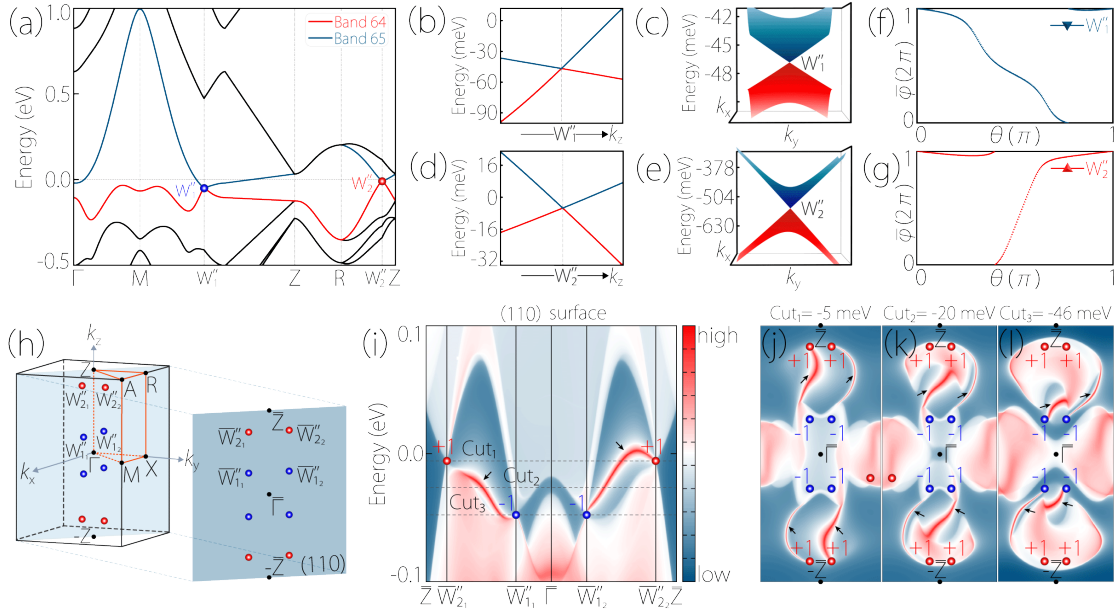

**Figure 6 | Symmetry-breaking degeneration into conventional Weyl fermions under a +2% uniaxial strain along the *a*-axis. a**, Bulk band structure of *r*-THRLN-$C_{32}$, showing that the explicit breaking of $C_4$ rotational symmetry shifts the band crossings ($W_1''$ and $W_2''$) to generic *k* points. **b-e**, Purely linear energy dispersions for both $W_1''$ and $W_2''$ along all three momentum directions ($k_x$, $k_y$, $k_z$), definitively characterizing them as conventional WPs. **f, g**, WCC evolution curves confirming the respective topological charges of -1 and +1. **h**, Spatial distribution of the eight WPs in the BZ and their corresponding (110) surface projections. **i**, Surface spectral function calculated on the (110) boundary. **j-l**, Isoenergy contours at -5 meV, -20 meV, and -46 meV, respectively, demonstrating the reversion to standard, unclosed open Fermi arcs connecting opposite-chirality WPs.

**Table I | Symmetry-allowed space groups (SGs) hosting precisely four double-Weyl points (DWPs, $C$-2) with and without spin-orbit coupling (SOC).** High-symmetry points (HSPs) and high-symmetry lines (HSLs) denote the momentum-space locations of the DWPs. PGs represents the corresponding crystallographic point group. IRRs refers to the irreducible (co)representations of the little group that characterize the $|C| = 2$ WPs.

| System | SGs (HSPs or HSLs) | PGs | IRRs |
|---|---|---|---|
| Four $C$-2 WPs | Without SOC | | |
| | 75 and 77 (Γ, M, Z, A, ΓZ, MA); 76 and 78 (Γ, M, ΓZ, MA); 79 (Γ, Z, P, ΓZ, ZV); 80 (Γ, Z, ΓZ, ZV) | $C_4$ | Γ, M, Z and A ({$R_2$,$R_4$}); P ({$R_1$,$R_2$}); ΓZ, MA and ZV ({$R_1$},{$R_3$};{$R_2$},{$R_4$}) |
| | 89 and 93 (Γ, M, Z, A, ΓZ, MA); 90 and 94 (Γ, Z, ΓZ); 91 and 95 (Γ, M, ΓZ, MA); 92 and 96 (ΓZ); 97 (Γ, Z, P, ΓZ, ZV); 98 (Γ, Z, ΓZ, ZV) | $D_4$ | Γ, M, Z and A ($R_5$); P ({$R_3$,$R_4$}); ΓZ, MA and ZV ({$R_1$},{$R_3$};{$R_2$},{$R_4$}) |
| | 168, 171 and 172 (Γ, A, ΓA); 169, 170 and 173 (ΓA) | $C_6$ | Γ and A ({$R_2$,$R_6$};{$R_3$,$R_5$}); ΓA ({$R_1$},{$R_3$};{$R_1$},{$R_5$};{$R_2$},{$R_4$}; {$R_2$},{$R_6$};{$R_3$},{$R_5$};{$R_4$},{$R_6$}) |
| | 177, 180 and 181 (Γ, A, ΓA); 178, 179 and 182 (ΓA) | $D_6$ | Γ and A ($R_5$;$R_6$); ΓA ({$R_1$},{$R_3$};{$R_1$},{$R_5$};{$R_2$},{$R_4$}; {$R_2$},{$R_6$};{$R_3$},{$R_5$};{$R_4$},{$R_6$}) |
| | With SOC | | |
| | 75-78 (ΓZ, MA); 79 (ΓZ, ZV); 80 (P, ΓZ, ZV) | $C_4$ | P ({$R_4$,$R_4$}); ΓZ, MA and ZV ({$R_2$},{$R_6$};{$R_4$},{$R_8$}) |
| | 89, 91, 93 and 95 (ΓZ, MA); 90, 92, 94 and 96 (ΓZ); 97 (ΓZ, ZV); 98 (P, ΓZ, ZV) | $D_4$ | P ({$R_1$,$R_4$}); ΓZ, MA and ZV ({$R_2$},{$R_6$};{$R_4$},{$R_8$}) |
| | 168-173 (ΓA) | $C_6$ | ΓA ({$R_2$},{$R_{10}$};{$R_2$},{$R_6$};{$R_4$},{$R_{12}$}; {$R_4$},{$R_8$};{$R_6$},{$R_{10}$};{$R_8$},{$R_{12}$}) |
| | 177-182 (ΓA) | $D_6$ | ΓA ({$R_2$},{$R_{10}$};{$R_2$},{$R_6$};{$R_4$},{$R_{12}$}; {$R_4$},{$R_8$};{$R_6$},{$R_{10}$};{$R_8$},{$R_{12}$}) |